\documentclass[,preprint,prl,amsmath,amssymb,aps,12pt]{revtex4}
\usepackage{appendix}
\usepackage{graphicx}
\usepackage{epstopdf}
\usepackage{color}
\usepackage[switch]{lineno}

\newcommand{\bea}{\begin{eqnarray}}
\newcommand{\eea}{\end{eqnarray}}

\begin{document}
\title{Influence of Micro-turbulence on Neoclassical Tearing Mode Onset}

\author{Tonghui Shi,$^1$ L. Wei,$^2$ H.H. Wang,$^1$ E. Li,$^1$ B. Shen,$^1$ J.P. Qian,$^1$ Y.M. Wang,$^1$ T. Zhang,$^1$ H.L. Zhao,$^1$ L. Zeng,$^1$ Y. Zhang,$^1$ H.Q. Liu,$^1$ Q. Ma,$^1$ D.L. Chen,$^1$ Z.P. Luo,$^1$ Y.Y. Li,$^1$ Z.C. Shen,$^1$ L.Q. Xu,$^1$ B. Zhang,$^1$ M.H. Li,$^1$ Z.X. Wang,$^2$ B.L. Ling,$^1$ X.Z. Gong,$^1$ Y. Sun$^{1,*}$ and B. Wan$^{1,\dag}$}

\affiliation{ $^1$Institute of Plasma Physics, Chinese Academy of Sciences, Hefei 230031, China
\\$^2$School of Physics, Dalian University of Technology, Dalian 116024, China}

\begin{abstract}
Direct evidence of micro-turbulence effect on the onset of neoclassical tearing mode (NTM) is reported for the first time in this letter. A puzzling positive correlation between critical width of seed island of NTM and normalized plasma pressure $\beta_p$ is first observed employing a novel method for clearly separating the processes of seed island and the onset of NTM in the EAST tokamak. Different from the methods developed before, the width of the seed island is well controlled by slowly ramping up the current in resonant magnetic perturbation coils. It is revealed that the positive correlation is mainly attributed to the enhancement of perpendicular transport by micro-turbulence, which overcomes the destabilizing effect of $\beta_p$ on the onset of NTM. Reduced magnetohydrodynamics (MHD) modeling well reproduced the two states of nonlinear bifurcations observed in this experiment by including the finite transport effect. This result provides a new route for understanding multi-scale interaction in plasma physics.
\end{abstract}
\maketitle
Magnetic island formed by tearing mode instability is a focus issue in space and laboratory plasmas, which provides an excellent medium to study bifurcation phenomena and multi-scale interaction between micro-turbulence and macro-instability in magnetized plasmas \cite{Magnetic recnection RMP 2010,Ida.K 2020}. In tokamak plasmas, Neo-classical Tearing Mode (NTM), namely nonlinear neoclassical pressure driven tearing mode, is one of the significant threats to a fusion reactor \cite{TFTR 1995 PRL}. It has been observed that small-scale micro-turbulence can be influenced by large-scale magnetohydrodynamics (MHD) instabilities \cite{Bardoczi PRL 2016,Sunpj NF 2018}, and in turn its enhanced cross-field transport can affect the stability of large magnetic island \cite{Bardoczi pop 2017a}. As a nonlinear MHD instability, NTM requires a large enough seed island to be triggered \cite{LaHaye 2006}. Plasma theory predicted small-scale micro-turbulence can affect the threshold width of seed island for triggering NTM \cite{Sanae PRL 2003,CaiHS NF 2019}. In this letter, we report for the first time a direct evidence of the influence of micro-turbulence on the onset of NTM by a new method to uncover the entire growth process of NTM, i.e. separating the seed island phase and the nonlinear triggering of NTM phase by well controlling the ramping rate of the seed island width using resonant magnetic perturbations (RMPs), and present a distinct dependency between critical width of seed island and normalized plasma pressure $\beta_p$.

Many achievements have been reached in NTM seed island research. In some experiments, NTM was found to be triggered by some MHD instabilities \cite{LaHaye 2006,NTM-HL2A,NTM-EAST,NTM-KSTAR}, in which its triggering process was very fast-about tens of milliseconds. In other experiments, using classical tearing mode and external magnetic perturbation as seed island and seed island driving force respectively \cite{TCV TM to NTM 2002,AUG 2019}, the seed island phase was demonstrated to sustain nearly $100ms$, and the bifurcation related to NTM triggering threshold was distinguished. Besides, marginal width of NTM magnetic island, indirectly reflecting the characteristics of NTM seed island, was obtained by ramping-down input power, presenting a proportional relation to ion banana orbit width \cite{LaHaye NF 2006 w_mar and rho_i}, which could be well explained by NTM theory including finite orbit width effect \cite{Poli PRL 2002,Imada PRL 2018,Wang Feng 2019}.
Although NTM theory, which contains a variety of effects \cite{Sauter 1997}, has successfully explained many observations \cite{LaHaye 2006}, uncertainties still remain in the physics describing small magnetic island, e.g. NTM seed island, because the growth process of NTM is difficult to be controlled and the plasma is unstable before NTMs saturated.

\begin{figure}[htbp]
\centering
\resizebox{12cm}{12cm}{\includegraphics{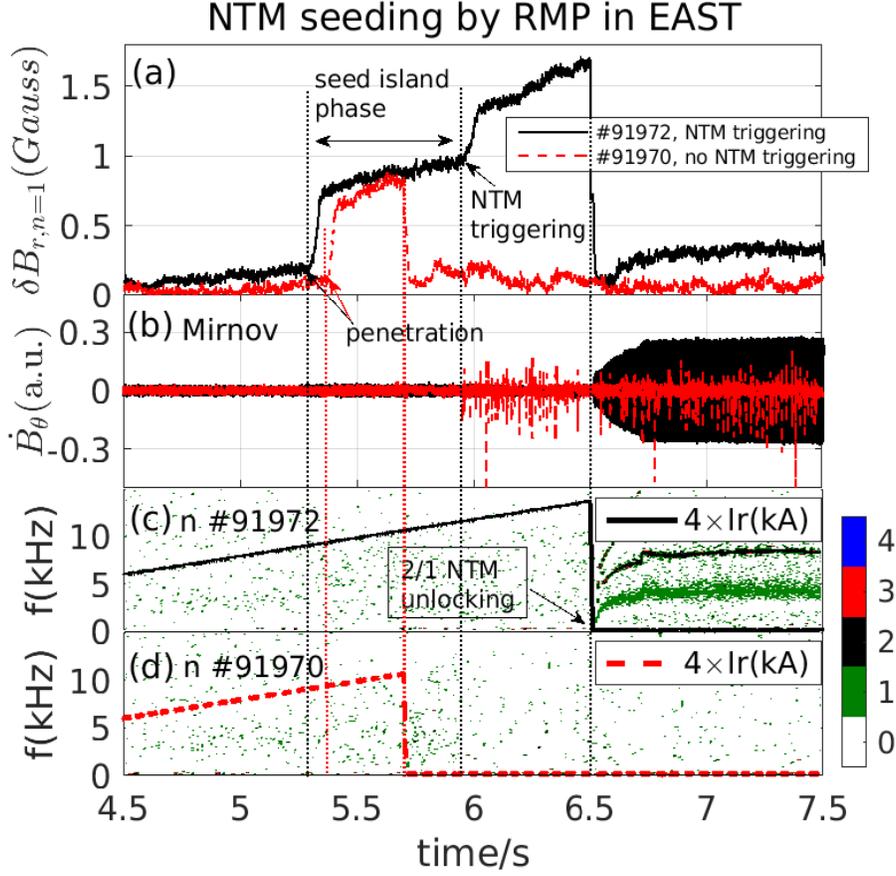}}
\caption{(color online). Temporal evolutions of
(a) $n=1$ perturbed magnetic field $\delta{B}_{r,n=1}$ measured by saddle loops,
(b) Mirnov signal $\dot{\delta{B}_\theta}$,
(c) and (d) spectra of Mirnov signals for toroidal mode number $n$ and RMP currents for shots $91970$ (red dashed lines) and $91972$ (black solid lines), respectively.}
\label{fig1}
\end{figure}

In this letter, we introduce a new method that the width of seed island is well controlled by slowly ramping up the current in RMP coils, seed island phase sustains hundreds of milliseconds, and the plasma is more stable before NTM triggering. EAST tokamak \cite{EAST 2019} has a flexible RMPs coil system consisting of $(2\times8)$ coils installed in the low side \cite{EAST RMP PRL 2016,EAST RMP NF 2017}. NTMs are often stable in EAST high confinement discharges due to low intrinsic error field \cite{EAST error field NF 2016}. All shots shown in this letter have been set as even coil connection between the upper and lower RMP coils, which can easily generate mode penetration in EAST \cite{EAST error field NF 2016}.

Fig. \ref{fig1} shows an example of the triggering of $m/n=2/1$ NTM by $n=1$ RMP in two shots $\sharp91970$ and $\sharp91972$, in which mode penetrations ($1th$-BF) are both observed when the RMP currents exceed some critical values, where magnetic island amplitude $\delta{B}_{r,n=1}$ (Fig. \ref{fig1}(a)) and poloidal and toroidal mode numbers $(m, n)$ are measured by $(5\times8)$ saddle coils installed in different toroidal and poloidal angles. The two shots have nearly same plasma parameters except the RMP current, i.e. magnetic safety factor $q_{95}\simeq5.4$, $\beta_N\simeq0.72$, $\langle{N}_{e0}\rangle\simeq2.6 (\times10^{19}/m^3)$ measured by POlarimeter-INTerferometer (POINT) system \cite{EAST POINT 2014}, $T_{i0}\simeq1.0kev$ measured by charge exchange recombination spectroscopy (CXRS) \cite{EAST CXRS 2014}, lower hybrid wave heating (LHW) \cite{EAST LHCD NF 2015} $P_{LHW}\simeq1.0MW$, and co-current neutral beam injection (NBI) \cite{EAST NBI 2015} $P_{NBI}=2.0MW$.
In shot $\sharp91972$ (Fig. \ref{fig1} black solid line), a bifurcation ($2nd$-BF) related to NTM triggering threshold ($t=5.9s$) is observed when RMP current grows to a critical value. Besides, the magnetic island unlocks when RMP current is turned off as shown in Mirnov signal (Fig. \ref{fig1}(b) black solid line) and its spectrum (Fig. \ref{fig1}(c)). For comparison, in shot $\sharp91970$ (Fig. \ref{fig1} red dashed line), if magnetic island width induced by RMP doesn't exceed some critical value, NTM can not be triggered and magnetic island decays immediately as RMP current is turned off (Fig. \ref{fig1}(a) and \ref{fig1}(b) red dashed lines and \ref{fig1}(d) spectrum). Therefore, by slowly ramping up the current in RMP coils, seed island width is well controlled, and the growth of seed island and the onset of NTM are clearly separated in this experiment.

\begin{figure}
\centering
\resizebox{12cm}{12cm}{\includegraphics{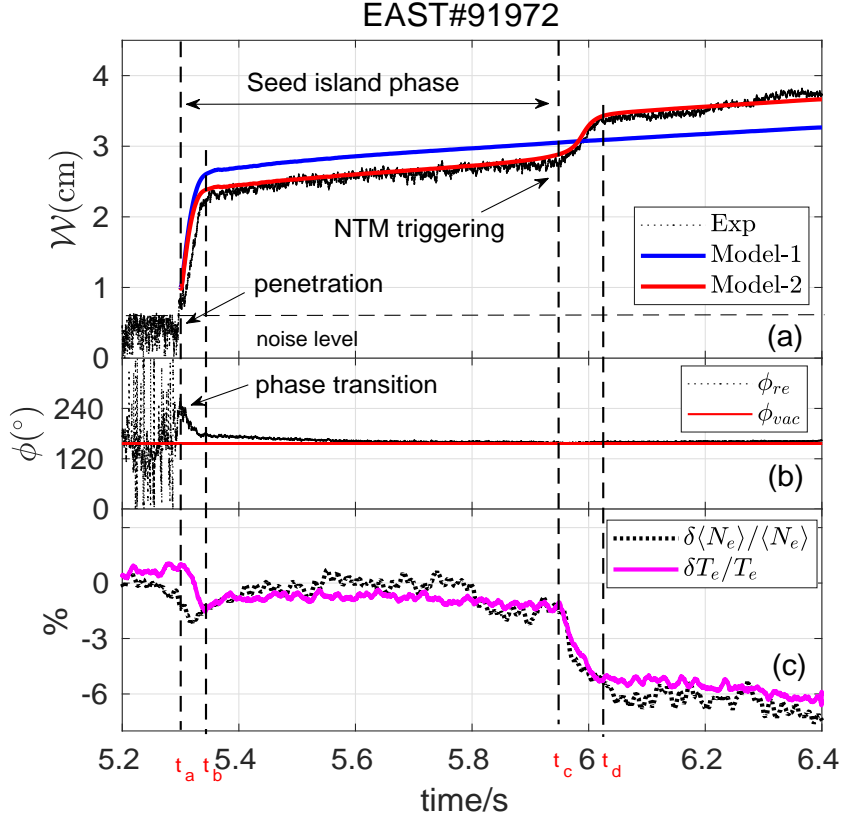}}
\caption{(color online). Temporal evolutions of (a) $2/1$ magnetic island width $\mathcal{W}$ from experiment (black doted line), the simulation results of mode-1 (blue line) and model-2 (red line) by using eq. \eqref{MRE},
(b) toroidal phase of $n=1$ perturbations in plasma response (black doted line) and vacuum field (red solid line), respectively,
(c) normalized variations of line averaged density (doted line) and perturbed electron temperature (solid line) near the $q=2$ surface.
}
\label{fig2}
\end{figure}

Applying this method, the dynamics of NTM triggering are clearly displayed. As shown in Fig. \ref{fig2}, the rapid growth of seed island width (Fig. \ref{fig2}(a) black dotted line) during $t_a<t<t_b$ is resulted from the fast-change of $\phi_{re}$ (Fig. \ref{fig2}(b) black dotted line), where plasma response phase $\phi_{res}$ and vacuum field phase $\phi_{vac}$ are measured by saddle coils, island width $\mathcal{W}\simeq{a}_{eq}\sqrt{\delta{B}_{r,n=1}}$, and $a_{eq}$ is related to plasma equilibrium parameters \cite{island formula PRL 2002}. The width of seed island during $t_b<t<t_c$ grows slowly and is well controlled by slowly ramping up the current in RMP coils. After phase transition (Fig. \ref{fig2}(b), $t>t_b$), the plasma response phase is nearly locked on vacuum field, even when $2nd$-BF ($t=t_c$) happens. The normalized perturbations near the $q=2$ surface, i.e. electron temperature $\delta{T_e}/{T_e}$ (Fig. \ref{fig2}(c) colored solid line) and electron density $\delta{\langle{N_e}\rangle}/\langle{N_e}\rangle$ (Fig. \ref{fig2}(c) black doted line), decline near $1th$-BF $(t=t_a)$ and $2nd$-BF $(t=t_c)$, where $T_e$ and $\langle{N_e}\rangle$ are measured by electron cyclotron emission (ECE) system \cite{EAST ECE 2018} and POINT respectively. But the declines near $2nd$-BF are sharper than that near $1th$-BF. This property of NTM triggering threshold has been disclosed in NTM theory and simulations \cite{Poli PRL 2002,Imada PRL 2018,Wang Feng 2019}. Using Eq. \eqref{MRE} with the parameters in table I, Model-2 (Fig. \ref{fig2}(a) red solid line) reconciles the experimental result (Fig. \ref{fig2}(a) black doted line), nevertheless, in Model-1 (Fig. \ref{fig2}(a) blue solid line), only one bifurcation is observed. In essence, the neoclassical driving force $\Delta_{bs}$ (Eq. \eqref{Delta}) becomes significant only when $\mathcal{W}>\mathcal{W}_{cri}$ in Model-2, which is consistent with experimental observations as shown in Fig. \ref{fig2}(c). Therefore, the dynamics of NTM triggering is clearly presented by this method, and the growth process of NTM is well simulated by MRE model as follows.

A Modified Rutherford Equation (MRE) \cite{Sauter 1997,Fitzpatrick NF 1993} with a switch $f_{bs}(\mathcal{W})$ about neoclassical effect (bootstrap current) is introduced as
\begin{equation}\label{MRE}
\frac{\tau_{R_{nc}}}{r_s}\frac{d\mathcal{W}}{dt}=r_s\Delta'+\Delta_{rmp}+\Delta_{bs},
\end{equation}
with
\begin{equation}\label{Delta}
\begin{split}
\left\{
\begin{aligned}
&\Delta_{rmp}=A_{rmp}\frac{\mathcal{W}_{vac}^2}{\mathcal{W}^2}\cos(\Delta\phi(t)), \\
&\Delta_{bs}= A_{bs}\beta_pf_{bs}\frac{{r_s}\mathcal{W}}{\mathcal{W}^{2}+\mathcal{W}_{cri}^{2}},
\end{aligned}
\right.
\end{split}
\end{equation}
where $f_{bs}(\mathcal{W})={1}/[1+(\mathcal{W}_{cri}/\mathcal{W})^{n_{bs}}]$, $\mathcal{A}_{rmp}$ and $\mathcal{A}_{bs}$ are related to plasma equilibrium parameters, $\tau_{R_{nc}}$ is the effective resistive time \cite{Sauter 1999}, vacuum magnetic island width $\mathcal{W}_{vac}$ calculated by MAPS code \cite{Sun MAPS code}, and $\Delta\phi(t)$ is phase difference. If setting $n_{bs}=0$, Eq. \eqref{MRE} becomes the well-known model (Model-1) \cite{Fitzpatrick finite transport 1995}. And if setting $n_{bs}\gg1$, Eq. \eqref{MRE} is Model-2, where $f(\mathcal{W})\simeq1$ when $\mathcal{W}>\mathcal{W}_{cri}$, and $f(\mathcal{W})\simeq0$ when $\mathcal{W}<\mathcal{W}_{cri}$.

\begin{table}
	\centering
	\caption{Modeling parameters in model-1 and model-2}
	\label{table1}
\begin{tabular}{p{40pt}p{40pt}p{40pt}p{40pt}p{40pt}p{28pt}}
		\hline\hline
		& & & & & \\[-12pt]
		$\tau_{R_{nc}}(s)$&$r_s\Delta'$&$A_{rmp}$&$A_{bs}\beta_p$&$\mathcal{W}_{cri}(m)$&$r_s(m)$ \\
		\hline
		& & & & & \\[-12pt]
	0.3&-2.61&6.78&0.24&0.03&0.23 \\
		\hline\hline
	\end{tabular}
\end{table}

\begin{figure}
\centering
\resizebox{11cm}{13cm}{\includegraphics{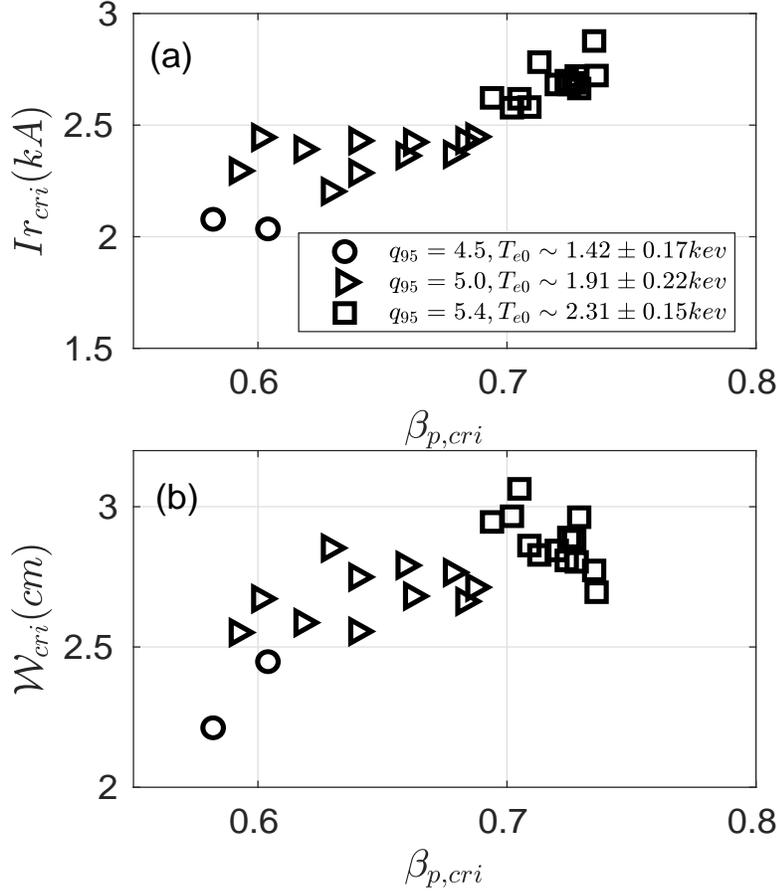}}
\caption{Dependencies of threshold RMP current $Ir_{cri}$ (a) and seed island critical width $\mathcal{W}_{cri}$ (b) on $\beta_{p,cri}$ at the NTM onset time.}
\label{fig3}
\end{figure}

Based on the statistical results, a puzzling positive correlation between seed island critical width and $\beta_p$ is discovered for the first time. As illustrated in Fig. \ref{fig3}, both RMP critical current value $Ir_{cri}$ and seed island critical width $\mathcal{W}_{cri}$ increase with $\beta_{p,cri}$, where the plasma parameters are similar except for $q_{95}\subset[4.5, 5.0, 5.4]$, i.e. $\langle{N_{e0}}\rangle=2.57\pm0.12(\times10^{19}/m^3)$, $T_{i0}=1.08\pm0.14kev$. As is well known, the fraction of bootstrap current is nearly proportional to plasma poloidal beta $\beta_p$. Then, dose it mean the bootstrap current has a stabilizing effect on NTM onset?

In fact, from NTM theory \cite{Sauter 1997}, neoclassical effect (bootstrap current) always plays a destabilizing role in NTM triggering. From a recent simulation \cite{Q Yu NF 2020}, bootstrap current shows a stabilizing effect on NTM onset because of particle pumping-in effect caused by RMP when magnetic island rotates in co-current direction before mode penetration. Nevertheless, in our case, seed island has been locked on vacuum field and island rotation remains nearly zero for a long time before $2nd$-BF happening. Moreover, curvature effect is very small in conventional high aspect ratio tokamaks and can't change the property of bootstrap current destabilization on NTM onset \cite{Sauter 1997}. And, we don't discuss neo-classical polarization current effect or finite orbit width effect \cite{Poli PRL 2002,Imada PRL 2018,Wang Feng 2019} because these shots have nearly same ion temperatures as shown in Fig. \ref{fig3}. In addition, in other NTM experiments \cite{NTM-COMPASSD,NTM-DIIID,NTM-EAST}, none has observed an obvious dependency between seed island threshold width and the friction of bootstrap current (or $\beta_p$), although bootstrap current is the main driving force of NTM.

\begin{figure}
\centering
\resizebox{12cm}{9cm}{\includegraphics{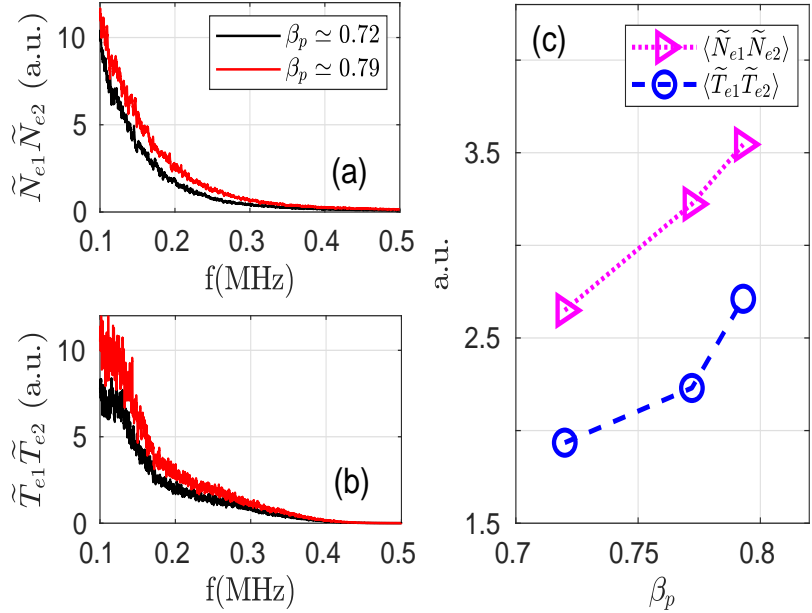}}
\caption{(color online). Local perturbations near the $q=2$ surface. Cross-powers spectra of (a) electron density $\widetilde{N}_{e1}\widetilde{N}_{e2}$ and (b) electron temperature $\widetilde{T}_{e1}\widetilde{T}_{e2}$.
(c) Dependencies of $\langle\widetilde{N}_{e1}\widetilde{N}_{e2}\rangle$ and $\langle\widetilde{T}_{e1}\widetilde{T}_{e2}\rangle$ on $\beta_p$, where $\langle\widetilde{f}_1\widetilde{f}_2\rangle_{f=T_e,N_e}$ presents the sum of cross-powers $\widetilde{f}_1\widetilde{f}_2$ in the $0.1MHz-0.5MHz$ frequency range.
}
\label{fig4}
\end{figure}
From the measurements of turbulence powers, it is found that the cross-field transport enhanced by micro-turbulence can affect the onset of NTM. As shown in Fig. \ref{fig4} (a) and (b), the amplitudes of cross-powers of electron density $\widetilde{N}_{e1}\widetilde{N}_{e2}$ and electron temperature $\widetilde{T}_{e1}\widetilde{T}_{e2}$ increase with $\beta_p$ in full $0.1MHz-0.5MHz$ frequency range, where fluctuations of $\widetilde{N}_{e1,2}$ and $\widetilde{T}_{e1,2}$ are measured by reflectometer \cite{EAST refectometry} and ECE respectively. These three similar shots have nearly same plasma equilibrium parameters except $\beta_p$ caused by different LHW heating powers. And the fluctuated amplitudes $\langle\widetilde{T}_{e1}\widetilde{T}_{e2}\rangle$ and $\langle\widetilde{N}_{e1}\widetilde{N}_{e2}\rangle$ have positive dependency on $\beta_p$ (Fig. \ref{fig4}(c)), which are associated with broadband low-$k$ turbulence powers and calculated by nearest neighbor cross-powers just before $2nd$-BF happening in the $0.1MHz-0.5MHz$ frequency range.
 Besides, in these shots (Figs. \ref{fig3} and \ref{fig4}), the increase of $\beta_p$ mainly arises from the increase of electron temperature caused by LHW heating. Consequently, the enhancement of turbulence power may be related to the gradient of electron temperature, and thus the cross-field transport enhanced by micro-turbulence plays a stabilizing role in the onset of NTM.

\begin{figure}
\centering
\resizebox{12cm}{10cm}{\includegraphics{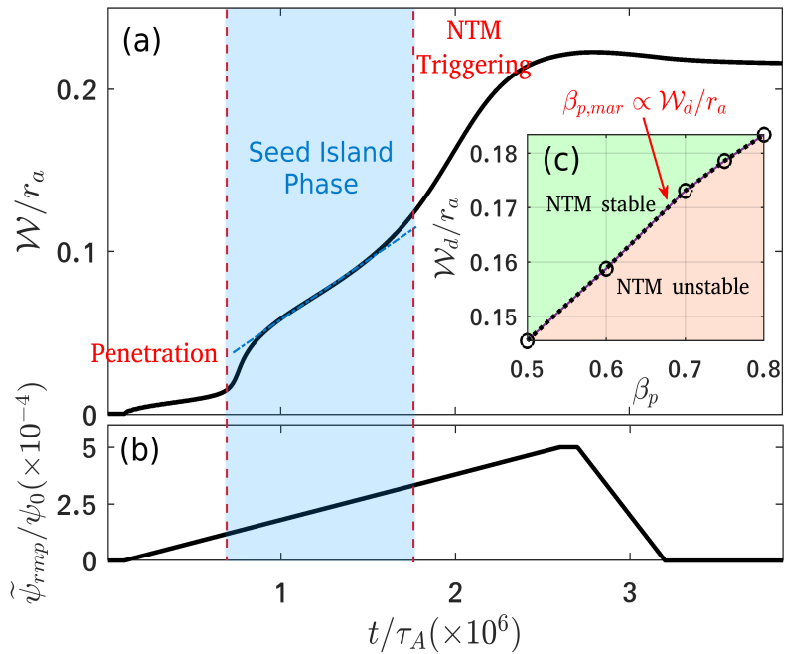}}
\caption{(color online).
 Temporal evolutions of (a) normalized magnetic island width,
(b) input normalized RMP amplitude.
(c) NTM marginally stable boundary: $\beta_{p,mar}\propto\mathcal{W}_d/r_a$. }
\label{fig5}
\end{figure}

These observations are well explained by NTM theory with the finite transport effect.
Using a reduced MHD code including the finite transport effect \cite{L Wei NF 2016}, the two states of nonlinear bifurcations in this experiment (Fig. \ref{fig1}) are well reproduced as shown in Fig. \ref{fig5} (a) and (b). When the plasma is close to marginally stable state of NTM (i.e. the fraction of bootstrap current is not large enough), this positive correlation between seed island critical width and $\beta_p$ is also simulated as illustrated in Fig. \ref{fig5}(c), in which the stable and unstable regions of NTM are clearly separated by the marginal stable boundary of NTM, i.e. $\beta_{p,mar}\propto{\mathcal{W}_{d}}/r_a$, where the scale length of finite transport is calculated by a formula $\mathcal{W}_d=[5.1r_s/(\epsilon_sns_s)^{1/2}](\chi_\perp/\chi_\parallel)^{1/4}$ \cite{Fitzpatrick finite transport 1995}, $\epsilon_s=r_s/R$ and local magnetic shear $s_s=rq'/q|_{r=r_s}$.
And, $\chi_\perp/\chi_\parallel$ and $\beta_{p,mar}$ are the output of this reduced MHD code.
In addition, applying MRE model Eq. \eqref{MRE}, we also obtain a proportional dependency, i.e. $\beta_{p,cri}\sim\beta_{p,mar}=-(r_s\Delta'/\mathcal{A}_{bs})(\mathcal{W}_{cri}/r_s)$ \cite{NTM-COMPASSD}, where $\mathcal{W}_{cri}\sim\mathcal{W}_d$ related to perpendicular transport strength enhanced by micro-turbulence.
Therefore, when plasma is close to the marginal state of NTM, micro-turbulence enhanced cross-field transport overcomes the destabilization of bootstrap current and shows a distinct stabilizing effect on NTM onset.

In summary, direct evidence of the influence of micro-turbulence on the onset of NTM is reported for the first time in the EAST tokamak. Applying a novel method, a positive dependency between critical width of seed island of NTM and plasma poloidal beta is first discovered. Different from the methods developed before, the process of seed island and the onset of NTM is distinguished more clearly, and the width of the seed island is well controlled by slowly ramping up the current in RMPs. From the analysis of fluctuated amplitudes associated with turbulence power, it is revealed that the positive dependency is mainly resulted from the enhancement of cross-field transport by micro-turbulence, which overcomes the destabilizing effect of $\beta_p$ and shows a stabilizing effect on the onset of NTM. Finally, using a reduced MHD model including the finite transport effect, the two bifurcations related to mode penetration and NTM triggering in this experiment are well reproduced, and the stabilizing effect of perpendicular transport enhanced by micro-turbulence on the NTM onset is also simulated. In brief, this method offers remarkable potentials for studying magnetic island dynamics and multi-scale interaction between micro-turbulence and small-scale magnetic island.

One of authors, Tonghui Shi, would like to thank Q. Yu (Max-Planck-Institut f\"{u}r Plasmaphysik, Germany) for helpful discussion. This work is supported by the National Key R$\&$D Program of China (Grants No: 2017YFE0301100, 2018YFE0302100, 2019YFE03010002), and the National Natural Science Foundation of China (Grants No: 11905250 and 11875292).

$^*$ywsun@ipp.ac.cn, $^\dag$bnwan@ipp.ac.cn

\end{document}